\documentclass[]{raa}            
\usepackage{graphicx,times}             
\usepackage{natbib}
\usepackage{amssymb,amsmath}
\usepackage{pdflscape}
\usepackage{color}
\bibpunct{(}{)}{;}{a}{}{,}

\usepackage[a4paper=true,pagebackref=true]{hyperref}
\hypersetup{colorlinks = true, linkcolor = green, anchorcolor = red, citecolor = blue, filecolor = red, pagecolor = red, urlcolor = red}

\newcommand{\fv}{$F_{\rm var}$}
\newcommand{\gr}{$\gamma$-ray}

\newcommand{\fermi}{\textit{Fermi}}

\begin{document}


   \title{{\it Fermi} Blazars in the Zwicky Transient Facility Survey: Properties of Large Optical Variations}

    \volnopage{Vol.0 (20xx) No.0, 000--000}      
   \setcounter{page}{1}          

   \author{Si-Si Sun
      \inst{1}
   \and Zhong-Xiang Wang
      \inst{1,2}
   \and Shun-Hao Ji
      \inst{1}
   }


   \institute{Department of Astronomy, School of Physics and Astronomy, Yunnan University, Kunming, Yunnan, 650091, China 
   {\it wangzx20@ynu.edu.cn}\\
        \and
	Shanghai Astronomical Observatory, Chinese Academy of Sciences, 80
Nandan Road, Shanghai 200030, China\\
      {\small Received~~20xx month day; accepted~~20xx~~month day}}

\abstract{ 
We analyze the optical light-curve data, obtained with the Zwicky Transient
Facility (ZTF) survey, for 47 \gr\ blazars monitored by the Large Area
Telescope onboard {\it the Fermi Gamma-ray Space Telescope (Fermi)}.
These 47 sources are selected because they are among the \fermi\ blazars
with the largest optical variations in the ZTF data. Two color-magnitude 
variation patterns
are seen in them, one being redder to stable when brighter (RSWB;
in 31 sources) and the other being stable when brighter (in 16 sources). 
The patterns fit with the results recently reported in several similar studies
with different data.
Moreover, we find that the colors in the stable state 
of the sources share similar values, which (after corrected for the Galactic 
extinction) of most sources are in a range of 0.4--0.55. This feature could
be intrinsic and may be applied in, for example, the study of intragalactic
medium. We also determine the turning points for the sources showing the
RSWB pattern, after which the color changes saturate and become stable. We find
a correlation between optical fluxes and \gr\ fluxes at the turning points.
The physical implications of the correlation remain to be investigated, probably
better with a sample of high-quality \gr\ flux measurements.
}

   \authorrunning{Sun et al.}            
   \titlerunning{Properties of large optical variations of blazars}  

   \maketitle

\section{Introduction} \label{sec:intro}

As a subclass of the active galactic nuclei (AGNs) that are powered by 
accretion of matter onto central supermassive black holes of 
galaxies \citep{urr+95,Ghi+98}, blazars are those with a jet pointing
close to our line of sight. They thus exhibit highly variable radiation 
across the whole electromagnetic spectrum due to the Doppler beaming 
effect (e.g., \citealt{Hov+09,Hov+19}),
since their emissions we observe are mostly from the jets.
Blazars have a double-hump structure of broadband spectral energy distributions 
(SEDs), with the low-energy hump peaking at frequencies from radio to 
ultraviolet/X-rays and the high-energy hump at energies from hard X-ray to \gr.
Generally, a so-called leptonic scenario is often considered for the emissions
of blazars, in which the low-energy hump arises from
synchrotron radiation of relativistic electrons accelerated in a jet, 
and the high-energy hump from inverse Compton Scattering (ICS) of photons 
by the same population of the electrons, where the seed photons are provided
by either the jet (e.g.,\citealt{Mar+92,Blo+96})
or/and the other AGN components (e.g., broad-line region,
dusty torus, and accretion disk; \citealt{Sik+94,Der+94,B00,Ghi+09}).
\begin{figure}
\centering
\includegraphics[width=0.44\textwidth]{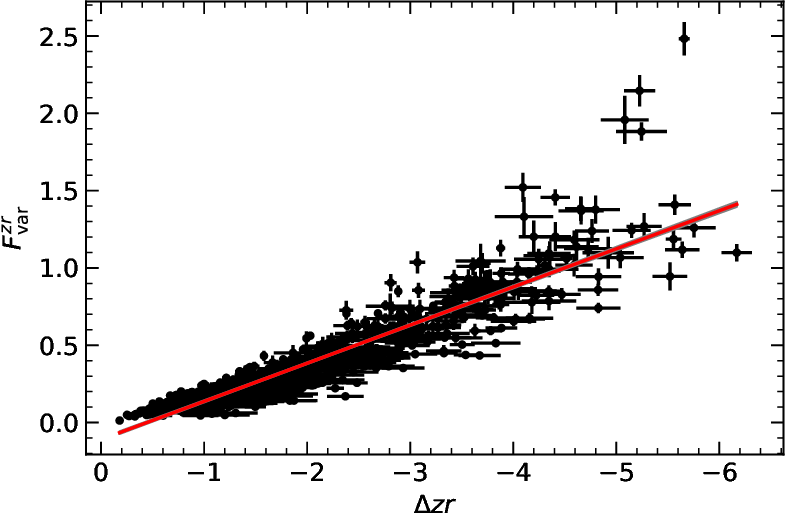}
\includegraphics[width=0.4\textwidth]{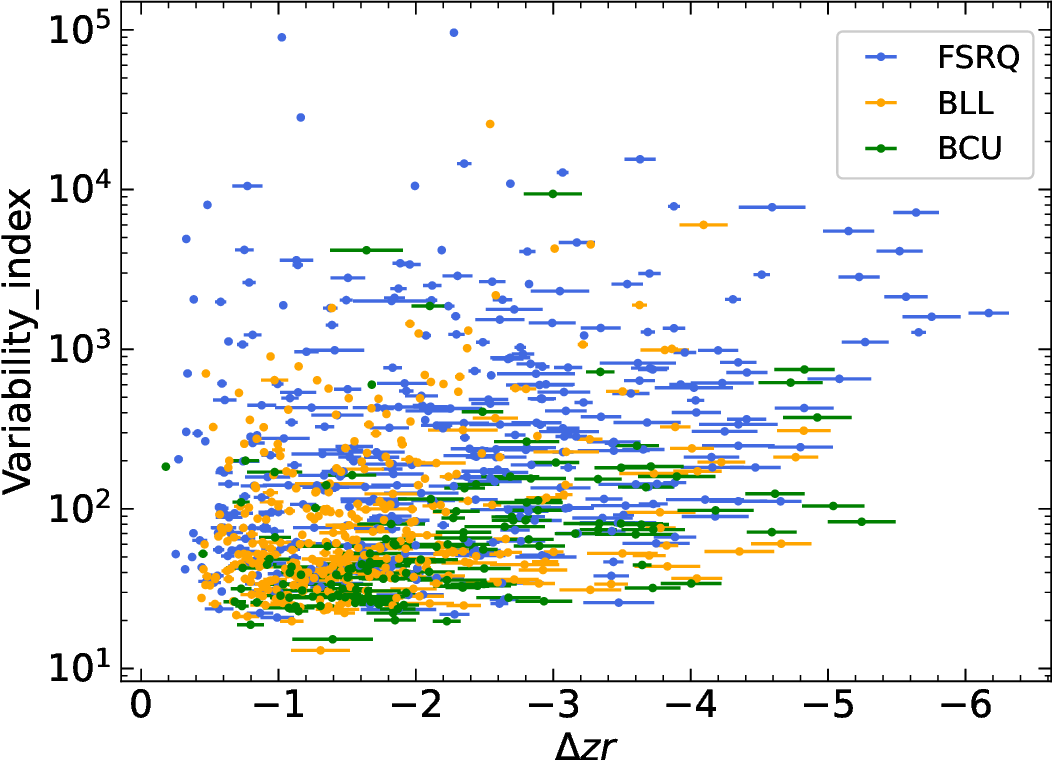}
	\caption{{\it Left:} $\Delta zr$ and $F_{\rm var}$ in the $zr$ band
	calculated for 810 blazars, which can be described with a fit of 
	$F_{\rm var} = -0.25 \Delta zr - 0.11$ (red line).
	{\it Right:} \gr\ variability versus $\Delta zr$ for the 810 blazars.
	No obvious variability correlation between the two bands is seen.}
 \label{fig:var}
\end{figure}

Since the launch of {\it the Fermi Gamma-ray Space Telescope (Fermi)} in 2008
\citep{Atw+09}, blazars have been fully observed at $\gamma$-rays by 
the Large Area Telescope (LAT) onboard \fermi\ in energy range of 
approximately from 100\,MeV to 500\,GeV. Not only
have nearly 4000 blazars been detected \citep{Ballet+23}, but their emissions 
also have been
monitored since 2008. Thus, \fermi-LAT blazars constitute a valuable sample
for different studies. In particular, \gr\ light curves binned on three 
different timescales for more than 1400 blazars 
are being provided by the \fermi-LAT light curve repository 
(LCR; \citealt{flcr}). Flux and spectral variation information for 
these blazars can be extracted for studies.

Variability studies of blazars help probe their properties and related
radiation mechanisms, jet
properties, and emission regions (see, e.g., \citealt{abd+10c,bot19}).
For example, the short, intraday variations are used to constrain
the sizes of the emission regions, while the ultra-short ones (on timescales
of a few minutes) may further indicate the detailed emission scenarios
(e.g., \citealt{abk17} and references therein). A more common type of 
variations seen in blazars may be described with a stochastic process 
with time scales of $\sim$10--100\,days 
(e.g., \citealt{zyz23} and references therein), accompanied with
(large) flares (e.g., \citealt{Hay+15}). Study of the variations probe
the jets' activity and their potential connection with accretion disks,
and multi-band monitoring of flares provide clues for emission scenarios
and locations of emission regions through analyses of
broadband SEDs and variation correlations between different 
bands (e.g., \citealt{Hay+15}; \citealt{kra+16}; \citealt{Coh+14}; 
\citealt{Lio+18}; \citealt{BG21}; \citealt{swx23}).

Among variability studies of blazars, one is to investigate the 
color-magnitude changes at optical bands 
(e.g., \citealt{Gu+06,Ote+22,Zha+22,Mc+24}). The investigations reveal
the optical spectral changes during activities of blazars and possibly
indicate the emission contribution from accretion 
disks (e.g., \citealt{rai+08,Isl+17,fan+18}). Such studies can be widely
carried out,
with the advent of large surveys of optical transients. A rich amount of
data at decent cadences covering large sky areas are provided by
these surveys. In this
work, we used the data from the Zwicky Transient Facility (ZTF;
\citealt{Bel+19}) survey to study the optical variations of \fermi\ blazars.
Interesting color-magnitude variation patterns from a sample of 47 blazars
were found. In addition, simultaneous \gr\ and X-ray data were analyzed
for helping understand the optical variations. We report the results in 
this paper. 

Below in Section~\ref{sec:dt}, we first describe the data we used and 
the target selection procedure, where 47 blazars 
with the largest $zr$-band variations in the ZTF data were selected.
In Section~\ref{sec:ana}, we describe the analyses we conducted 
for the 47 blazars using the ZTF optical and \fermi-LAT \gr\ light-curve data;
available {\it Neil Gehrels Swift Observatory (Swift)} X-ray data were also 
analyzed.  We present the results and discuss them in Section~\ref{sec:rd}.

\begin{figure}{}
\centering
\includegraphics[width=0.5\textwidth]{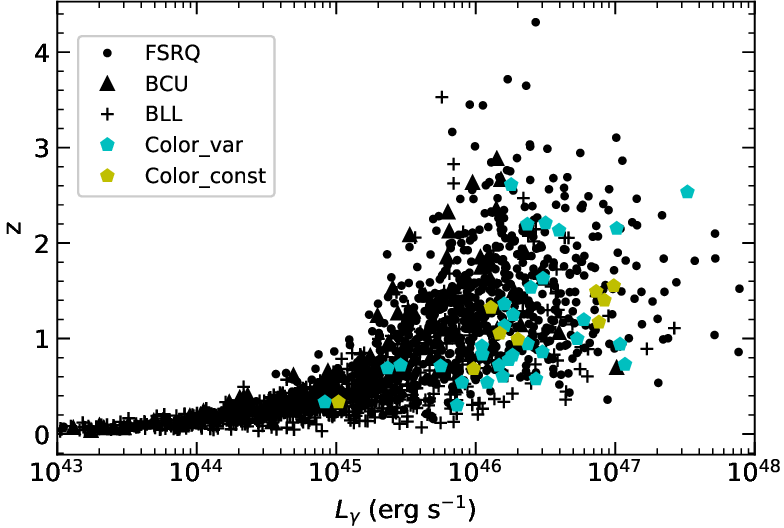}
  \caption{Redshifts versus \gr\ luminosities of \fermi-LAT detected blazars.
Among our 47 blazar targets, 40 have redshifts, and they are marked with
	pentagons.}
 \label{fig:lum}
\end{figure}

\section{Data and Target Selection}
\label{sec:dt}

\subsection{\fermi-LAT \gr\ and ZTF optical light-curve data}

\fermi-LAT $\gamma-$ray light curves from its LCR \citep{flcr} were used in 
our analysis. 
This public database provides light curves for \fermi-detected variable sources 
binned on timescales of 3-day, 7-day (weekly), and 30-day (monthly).
There were 1414 blazars among the \fermi-LAT source catalog 
that have optical counterparts with given coordinates 
and are included in \fermi-LAT LCR\footnote{https://fermi.gsfc.nasa.gov/ssc/data/access/lat/LightCurveRepository/}. 
We downloaded the monthly light curves for these 1414 blazars,
and used the flux data 
points with test statistic (TS) values greater than 4 (i.e., detection 
significances of the data points were greater than 2$\sigma$). 

We obtained the coordinates of 1414 blazars from the \fermi\ LAT source
catalog \citep{Ballet+23}, and searched for their magnitude data 
in the ZTF Data Release 20 with the provided API\footnote{\url{https://irsa.ipac.caltech.edu/docs/program_interface/ztf_lightcurve_api.html}}. 
The ZTF survey can provide magnitude data points at a cadence of 
as high as $\sim$2\,day for sources in its field.
To ensure high-quality of the light-curve data, we selected data points with 
catflags $=0$ and chi $<4$. We found 1403 blazars covered by the ZTF survey.
There are three bands, $zg$, $zr$, and $zi$, in the ZTF survey, and
we only used data at $zg$ and $zr$ bands since there often are limited data 
at $zi$ band.  Given our purpose of studying optical variations, which 
requires as many data points as possible for individual sources, we 
excluded those
sources with the number of data points less than 100 at $zr$ band.
After the exclusion, there were 810 blazars. The time period of the ZTF data
was approximately MJD~58190--60200.

\begin{table}
	\caption{Information for 47 blazar targets}
\centering
	\resizebox{\textwidth}{!}{
\begin{tabular}{cccccccc}
\hline
	4FGL & Name & Type & $z$ & L$_{\gamma}/10^{46}$ & $E(B-V)$ & $zr$ & $F_{\gamma}/10^{-8}$ \\ 
	 & & & & (erg\,s$^{-1}$) & (mag) & (mag) & (ph\,cm$^{-1}$\,s$^{-1}$)\\
\hline
J0022.5+0608 & PKS 0019+058         & BLL  & NULL   & NULL & 0.0197 & NULL   & NULL \\
J0028.5+2001 & TXS 0025+197         & FSRQ & 1.552  & 9.76 & 0.0794 & NULL   & NULL                 \\
J0038.2$-$2459 & PKS 0035$-$252  & FSRQ & 1.196   & 5.95 & 0.0129 & $19.39\pm0.13$ & $3.9\pm1.2$ \\
J0050.4$-$0452 & PKS 0047$-$051         & FSRQ & 0.92   & 1.11 & 0.0415 & $19.32\pm0.13$ & $2.2\pm1.3$ \\
J0107.4+0334 & PMN J0107+0333       & BLL  & NULL   & NULL & 0.0202 & NULL   & NULL                 \\
J0112.8+3208 & 4C +31.03            & FSRQ & 0.603  & 1.56 & 0.0491 & $15.33\pm0.14$ & $32.0\pm2.2$ \\
J0137.0+4751 & OC 457               & FSRQ & 0.859  & 3.00 & 0.1294 & $18.18\pm0.15$  & $12.7\pm1.8$ \\
J0238.6+1637 & PKS 0235+164         & BLL  & 0.94   & 10.8 & 0.0694 & $18.96\pm0.15$  & $5.0\pm3.0$ \\
J0401.7+2112 & TXS 0358+210         & FSRQ & 0.834  & 1.12 & 0.2122 & $19.44\pm0.12$ & $6.1\pm2.2$ \\
J0401.9$-$2034 & PMN J0401$-$2034       & BCU  & NULL   & NULL & 0.0083 & NULL   & NULL                 \\
J0407.5+0741 & TXS 0404+075         & BLL  & 1.13   & 1.61 & 0.2267 & $19.24\pm0.53$  & $2.2\pm1.5$ \\
J0418.1$-$0252 & PKS B0415$-$029        & BCU  & NULL   & NULL & 0.0279 & $20.48\pm0.19$ & $2.0\pm1.9$  \\
J0449.1+1121 & PKS 0446+11          & FSRQ & 2.153  & 10.2 & 0.4255 & $18.39\pm0.59$ & $21.3\pm2.5$ \\
J0634.9$-$2335 & PMN J0634$-$2335      & FSRQ & 1.535  & 2.48 & 0.0826 & NULL & NULL \\
J0713.8+1935 & MG2 J071354+1934     & FSRQ & 0.54   & 0.796 & 0.0963 & $18.57\pm0.13$ & $6.1\pm2.9$ \\
J0742.6+5443 & GB6 J0742+5444       & FSRQ & 0.72   & 1.47 & 0.0361 & $18.32\pm0.13$ & $1.8\pm1.2$  \\
J0809.5+5341 & 87GB 080551.6+535010 & FSRQ & 2.133  & 3.96 & 0.0363 & $19.63\pm0.46$ & $2.2\pm1.2$ \\
J1033.9+6050 & S4 1030+61           & FSRQ & 1.401  & 8.36 & 0.0081 & NULL   & NULL                 \\
J1040.5+0617 & GB6 J1040+0617       & BLL  & NULL   & NULL & 0.0224 & NULL   & NULL                 \\
J1049.8+1429 & MG1 J104945+1429     & BCU  & 1.63   & 3.03 & 0.0232 & $18.41\pm0.52$ & $9.9\pm2.4$  \\
J1159.5+2914 & Ton 599              & FSRQ & 0.729  & 11.7 & 0.0171 & $16.22\pm0.10$ & $13.9\pm1.6$ \\
J1251.3$-$0201 & TXS 1248$-$017         & BCU  & 0.335  & 0.828 & 0.0204 & NULL   & NULL                \\
J1303.0+2434 & MG2 J130304+2434     & BLL  & 0.993  & 2.0  & 0.0151 & NULL   & NULL                 \\
J1308.5+3547 & 5C 12.291            & FSRQ & 1.055  & 1.48 & 0.0079 & NULL   & NULL                 \\
J1310.5+3221 & OP 313               & FSRQ & 0.997  & 5.34 & 0.0115 & $17.78\pm1.11$  & $4.8\pm1.3$ \\
J1312.8$-$0425 & PKS B1310$-$041        & FSRQ & 0.8249 & 1.84 & 0.0282 & $18.57\pm0.12$  & $10.3\pm2.5$  \\
J1321.1+2216 & TXS 1318+225         & FSRQ & 0.943  & 2.36 & 0.0139 & $18.08\pm0.22$ & $7.2\pm1.8$ \\
J1332.6$-$1256 & PMN J1332$-$1256       & FSRQ & 1.492  & 7.30  & 0.0477 & NULL   & NULL                 \\
J1333.7+5056 & CLASS J13333+5057    & FSRQ & 1.362  & 1.60 & 0.0104 & $19.58\pm0.44$  & $1.54\pm0.91$ \\
J1337.6$-$1257 & PKS 1335$-$127         & FSRQ & 0.539  & 1.21 & 0.0658 & $16.94\pm0.42$  & $7.4\pm2.3$ \\
J1345.5+4453 & B3 1343+451          & FSRQ & 2.534  & 32.9 & 0.0197 & $19.13\pm0.33$ & $13.1\pm1.3$ \\
J1350.8+3033 & B2 1348+30B          & FSRQ & 0.7115 & 0.56 & 0.0132 & $17.52\pm0.06$ & $10.6\pm1.5$ \\
J1450.4+0910 & TXS 1448+093         & FSRQ & 2.611  & 1.79 & 0.0247 & $19.61\pm0.17$ & $2.0\pm1.4$ \\
J1453.5+3505 & MG2 J145315+3506     & FSRQ & 0.721  & 0.289 & 0.0129 & $20.46\pm0.10$ & $1.4\pm1.0$ \\
J1539.6+2743 & MG2 J153938+2744     & FSRQ & 2.196  & 2.34 & 0.0251 & $20.24\pm0.18$ & $0.81\pm0.68$  \\
J1639.2+4129 & MG4 J163918+4127     & FSRQ & 0.691  & 0.233 & 0.0061 & $18.15\pm0.21$ & $1.46\pm0.90$  \\
J1700.0+6830 & TXS 1700+685         & FSRQ & 0.301  & 0.735 & 0.0391 & $16.83\pm0.21$ & $15.7\pm1.9$ \\
J1841.0+6115 & 87GB 184000.4+611120 & BCU  & NULL   & NULL  & 0.0356 & $20.22\pm0.06$ & $3.3\pm1.7$ \\
J1852.4+4856 & S4 1851+48           & FSRQ & 1.25   & 1.85 & 0.0465 & $20.49\pm0.13$  & $1.52\pm0.98$ \\
J1954.6$-$1122 & TXS 1951$-$115         & BLL  & 0.683  & 0.966 & 0.1353 & NULL   & NULL             \\
J2007.2+6607 & TXS 2007+659         & FSRQ & 1.325  & 1.28 & 0.204 & NULL   & NULL                 \\
J2025.2+0317 & PKS 2022+031         & FSRQ & 2.21   & 3.16 & 0.0948 & $19.87\pm0.07$ & $2.6\pm1.8$ \\
J2244.2+4057 & TXS 2241+406         & FSRQ & 1.171  & 7.64 & 0.1455 & NULL   & NULL                 \\
J2301.0$-$0158 & PKS B2258$-$022        & FSRQ & 0.778  & 1.73 & 0.0521 & $17.17\pm0.28$ & $9.3\pm2.2$ \\
J2322.6$-$0735 & PMN J2322$-$0736       & BCU  & NULL   & NULL & 0.0310 & NULL   & NULL                 \\
J2326.2+0113 & SDSS J232625.63+011208.6 & BCU  & 0.335   & 0.083 & 0.0293 & NULL & NULL  \\
J2348.0$-$1630 & PKS 2345$-$16          & FSRQ & 0.576  & 2.70 & 0.0224 & $16.58\pm0.15$ & $6.6\pm1.7$  \\
\hline
\end{tabular}
}
	\tablecomments{0.86\textwidth}{Column 7 gives the magnitude of 
the $zr$ band at the turning point, Column 8 gives the corresponding
	$\gamma$-ray flux (in units of photon cm$^{-2}$\,s$^{-1}$), and
	NULL indicates no data.}
\label{Tab1}
\end{table}

\subsection{Target selection}

We checked variations of the 810 blazars using the $zr$-band data.
First,
the variation amplitudes $\Delta zr$ were obtained by subtracting 
the maximum magnitude from the minimum magnitude of the $zr$ light curve of 
each blazar.
Secondly, we calculated the fractional variability amplitude (\fv;
e.g., \citealt{sch+19}) of each
light curve as the further check. The formula for calculating $F_{\rm var}$ for
a light curve of $N$ data points is as follows:
\begin{equation}
F_{\rm{var}} = \sqrt{ \frac{S^{2} -
\overline{\sigma_{\rm{err}}^{2}}}{\bar{x}^{2}}}, 
\end{equation}

\begin{equation}
err(F_{\rm{var}}) =\sqrt{ \left\{ \sqrt{\frac{1}{2N}} \cdot\frac{ \overline{\sigma_{\rm{err}}^{2}}
}{  \bar{x}^{2}F_{\rm{var}} }  \right\}^{2}
+
\left\{ \sqrt{\frac{\overline{\sigma_{\rm{err}}^{2}}}{N}}
\cdot\frac{1}{\bar{x}}  \right\}^{2}  },
\end{equation}

\begin{equation}
\overline{\sigma_{\rm{err}}^{2}} = \frac{1}{N}\sum_{i=1}^{N} \sigma_{{\rm
err}, i}^{2} ,
\end{equation}

where $\bar{x}$ and $S^2$ are the mean flux and total variance 
of a light curve, respectively, $\sigma_{err}$ is the error of a flux
measurement, ${\overline{\sigma_{\rm{err}}^{2}}}$ is the mean square
error, and $err(F_{\rm{var}})$ is the error of
$F_{\rm{var}}$.
In calculating \fv, $zr$ magnitudes were converted to flux densities (in units
of Jy); the ZTF magnitudes are in the AB photometric system \citep{Bel+19}. 

In the left panel of Figure~\ref{fig:var}, we show 
$F_{\rm var}$ versus $\Delta zr$ of the 810 blazars. 
A relationship of $F_{\rm var} = -0.25 \Delta zr - 0.11$ can be obtained
by fitting the data points. The p-value of the fitting is extremely
small ($\simeq 10^{-16}$), indicating the high significance of the 
relationship. This comparison shows that the variability
of the blazars is approximately proportional to $\Delta zr$. 
In addition, we also checked $\gamma-$ray variability versus
$\Delta zr$ for these 810 blazars, where we used Variability\_index given
in the \fermi-LAT source catalog \citep{Ballet+23} to represent \gr\ 
variability.
As can be seen in the right panel of Figure~\ref{fig:var}, there are
no obvious correlated variabilities between the \gr\ and optical band.

Based on the optical-variation plot shown in Figure~\ref{fig:var}, 
we may use $\Delta zr$ to select 
targets for simplicity.  We found that there were 52 sources 
with the $\Delta zr$ variation greater than 4\,mag (i.e., sources
with largest variations, right of $-4$\,mag in the left panel of 
Figure~\ref{fig:var}). We decided to focus on these largest-variation
sources.  However, as we further analyzed
the data for them, we realized that five of them had too few 
$zg$ data points simultaneous to the $zr$ data points
in our process of obtaining their color variations.
We excluded these five sources and had 47 blazars as the final targets.
Information for these targets is provided in Table~\ref{Tab1}.

Blazars are divided into two sub-types based on emission-line features
detected in their optical emissions \citep{Sti+91,urr+95}: 
Flat Spectrum Radio Quasars (FSRQs) that show emission lines with an equivalent 
width $>$5 \AA\ and BL Lacertae objects (BL Lacs) that have weak
or no emission lines. We checked our targets and provide their sub-type 
information in Table~\ref{Tab1}. Among them, 33 are FSRQs and 7 are BL Lacs,
while the remaining 7 are blazar candidates of uncertain type (BCUs; e.g.,
\citealt{Ballet+23}). In addition, blazars are also classified 
by the peak frequencies $\nu^{sy}_{pk}$ of their synchrotron 
humps \citep{Abd+10,Fan+16,Ghi+17}: 
Low Synchrotron Peaked (LSP) blazars ($10^{14}$\,Hz $>\nu^{sy}_{pk}$), 
Intermediate 
Synchrotron Peaked (ISP) blazars ($10^{15}$\,Hz $>\nu^{sy}_{pk}>10^{14}$\,Hz), 
and High Synchrotron Peaked (HSP) blazars ($\nu^{sy}_{pk}>10^{15}$ Hz). 
We note that all the targets belong to LSP.

Finally we checked the redshifts ($z$) and \gr\ luminosities ($L_{\gamma}$)
of the targets. Among them, 40 have given redshifts. We show them
in Figure~\ref{fig:lum} along with other \fermi-detected blazars.
As can be seen, most of them are among the luminous blazars with
$L_{\gamma}$ in $\sim 10^{46}$--$10^{48}$\,erg\,s$^{-1}$, and there are
also 8 of them with $L_{\gamma}$ in $\sim 10^{45}$--$10^{46}$\,erg\,s$^{-1}$.
The bias towards luminous blazars (or FSRQs) could be set by our use of 
the ZTF data,
which have 5$\sigma$ detection limits of 20.8\,mag in $zg$ and 20.6\,mag 
in $zr$ \citep{Mas+19}. Thus, relatively more luminous blazars would
be well covered by the ZTF survey.

\section{Analysis} 
\label{sec:ana}

\subsection{Color-magnitude analysis}
\label{sec:opt}

A $zg-zr$ color curve for each target was calculated, where 
$zg$ and $zr$ magnitudes obtained within one same day were considered 
as simultaneous measurements and used in the calculation. Color-magnitude 
diagrams
of $zg-zr$ versus $zr$ for the targets were thus obtained. There are three types
of color-magnitude variations shown by our targets.  The first type is curved
changes following the redder when brighter (RWB) pattern, and after the color
reaches a point, it stays around a constant line without obvious trending. 
Following \citet{Zha+22}, we call the latter
constant-color part as the stable state.
Figure~\ref{fig:cur} shows a good example of this type. For this type,
we used the following function to fit the color-magnitude variations:

\begin{figure}
\centering
\includegraphics[width=0.8\textwidth]{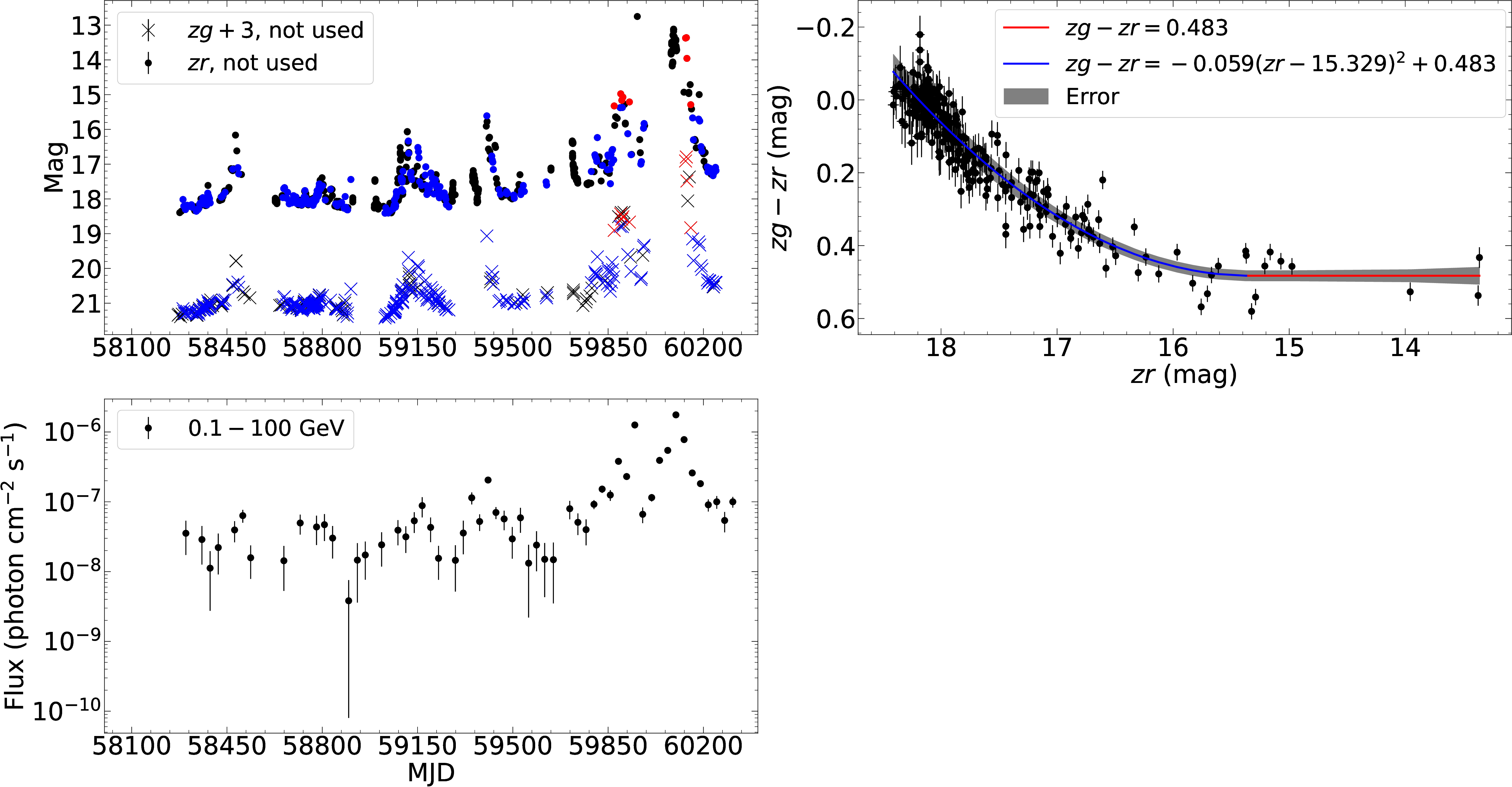}
	\caption{Light curve ({\it upper left}) and color-magnitude ({\it upper
	right}) diagrams of 4FGL J0112.8+3208. The curved and constant parts
	of the color changes are indicated by blue and red lines respectively
	in the color-magnitude diagram. The corresponding data points used to 
	construct the color-magnitude changes are respectively marked as blue
	and red in the light curves. The black data points ($zg$ or $zr$,
	not used) in the light curves are those do not have the same-day 
	measurements in the respective other band.
	The \gr\ light curve of the source during
	the ZTF time period is shown in the {\it lower left} panel.}
 \label{fig:cur}
\end{figure}

\begin{figure}
\centering
\includegraphics[width=0.8\textwidth]{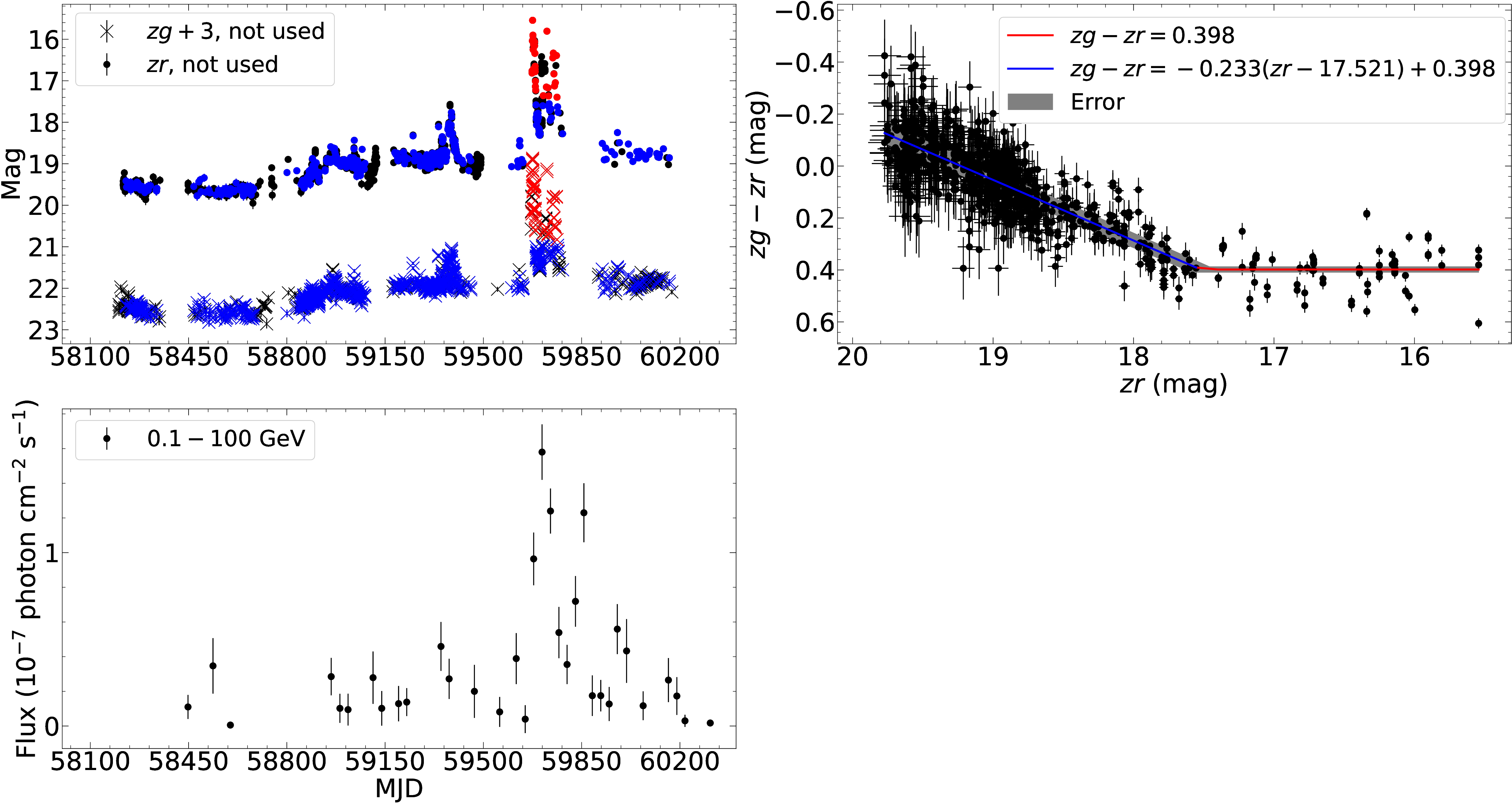}
	\caption{Same as Figure~\ref{fig:cur} for 4FGL J1350.8+3033. The RWB
	part is fitted with a straight line ({\it upper right} panel).}
 \label{fig:line}
\end{figure}

\begin{figure}
\centering
\includegraphics[width=0.8\textwidth]{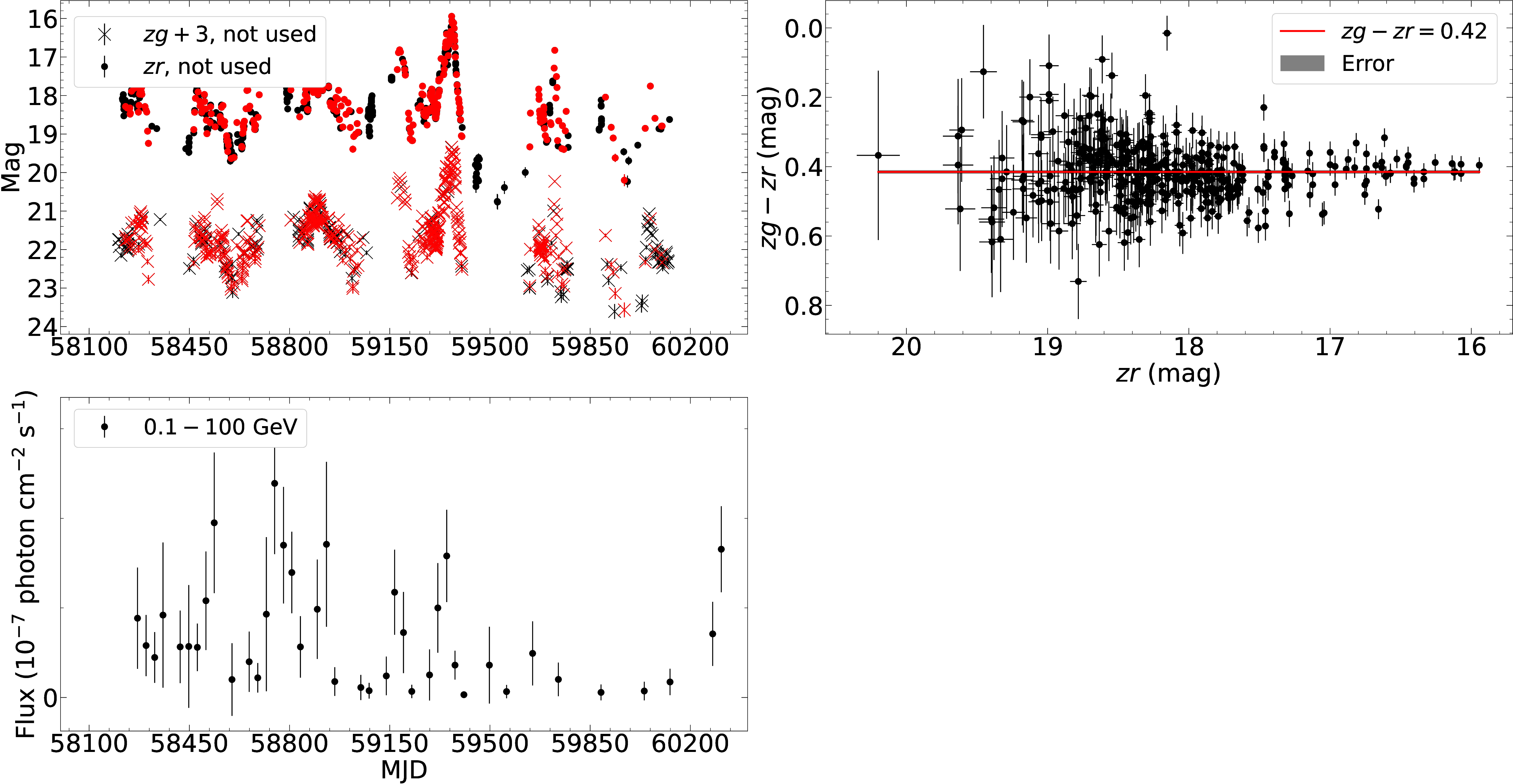}
	\caption{Same as Figure~\ref{fig:cur} for 4FGL J1303.0+2434. The color
	changes of this source can be described with a straight line
	({\it upper right} panel).}
 \label{fig:const}
\end{figure}

\begin{equation}
y = \left\{ \begin{array}
{r@{\quad,\quad}l}
k{(x-x_{\rm{b}})^2} + c
&
	{\rm if}\ x\geq x_{\rm{b}} \vspace{0.2 cm} 
\\
c 
&
	{\rm if}\ x < x_{\rm{b}} ,
\end{array} \right.
\end{equation}
where $x$ and $x_{\rm b}$ are the magnitude and the break point between
the curved changes and the constant $c$ line, respectively, and $y$ is the color
(note that we plot magnitudes and color values from large values to small
ones).
With this fitting (by using {\tt curve\_fit} in Python SciPy), 
we obtained both $x_{\rm b}$ and $c$.

The second type is similar to the first type, but the RWB changes appear linear.
Figure~\ref{fig:line} shows an example of this type of changes.
We fit the color-magnitude changes similarly to that of the first type, 
but used a straight line instead for the RWB changing part:

\begin{equation}
y = \left\{ \begin{array}
{r@{\quad,\quad}l}
k(x-x_{\rm{b}})+ c 
&
	{\rm if}\ x\geq x_{\rm{b}} \vspace{0.2 cm}
\\
c
&
	{\rm if}\ x < x_{\rm{b}} .
\end{array} \right.
\end{equation}
The $x_{\rm b}$ and $c$ values were obtained from the fitting. In total,
31 of our targets show these two types of changes, and
the first and second type is 16 and 15 respectively.

For the remaining 16 blazars, their color changes appear to be 
around a 
constant line without obvious RWB or other types of patterns seen. 
Figure~\ref{fig:const}
is an example of this type. For this third type, we fitted the color values
with a constant $c$. 

Besides the three example sources shown in 
Figures~\ref{fig:cur}--\ref{fig:const}, the
fits to the color-magnitude variations of the other 44 sources
are shown in Figures~\ref{fig:cm1} \& \ref{fig:cm2}.

\subsection{Determination of the \gr\ flux at $x_{\rm b}$}
\label{sec:gam}

We compared the \gr\ light curves with the optical ones for each target. 
The comparison indicates two different types of cases. One is 
that \gr\ flaring activity can be seen to correspond to optical flares (e.g., 
Figures~\ref{fig:cur} \& \ref{fig:line}) and the other is that there is
not clear flaring activity in \gr\ with respect to optical flare-like 
variations (e.g., Figure~\ref{fig:const}). Moreover, examining the optical
light curves and color changes, a common property can be drawn:
the constant-color part in the first two types of color changes 
(Section~\ref{sec:opt}) consists of
brighter magnitude data points contained in one or several flares.
This is clearly shown in Figures \ref{fig:cur} \& \ref{fig:line}
(red data points in the light curves).
Therefore it appears that there exists a turning point, 
$x_{\rm b}$ in our fitting, after which the color changes saturate and become
to be around a constant value (e.g., \citealt{fan+18}).

\begin{figure}{}
\centering
\includegraphics[width=0.5\textwidth]{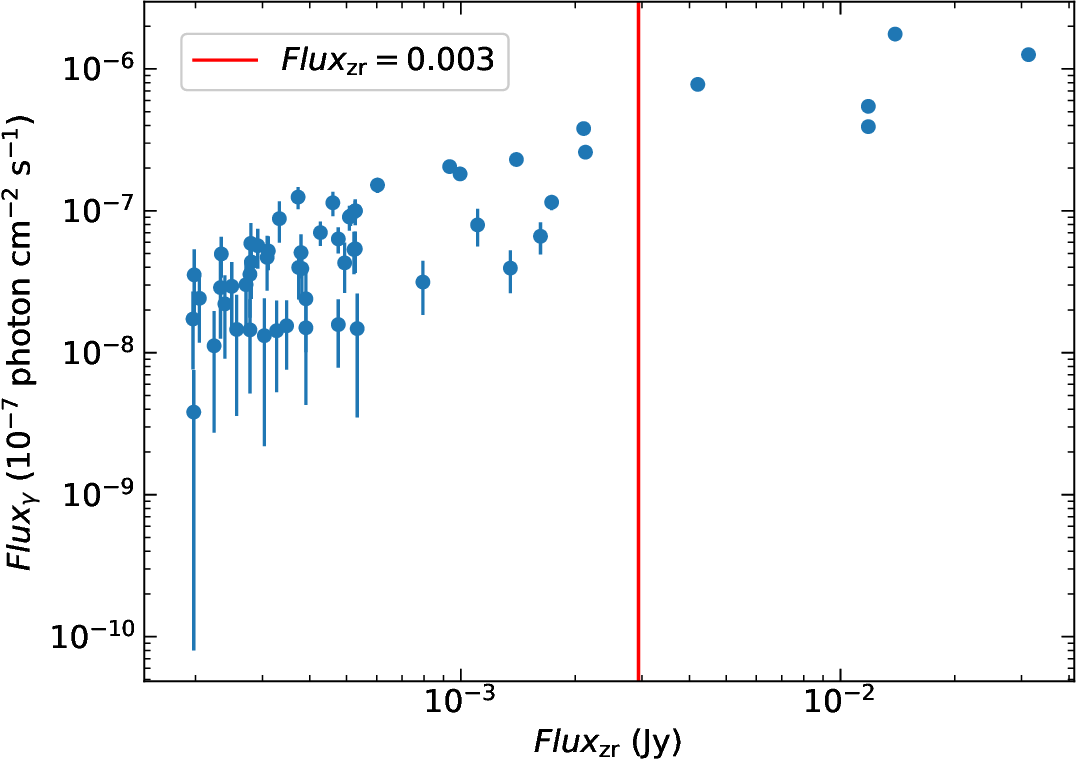}
  \caption{Example of determining the \gr\ flux at the turning point.
	\gr\ fluxes and corresponding $zr$ flux densities of 
	4FGL~J0112.8++3208 are shown. The red vertical line is 
	the $zr$-band flux density of $x_b$. The average of two 
	\gr\ fluxes closest
	in time to that of $x_b$ is taken as the value at the break point.}
 \label{fig:go}
\end{figure}

In order to connect the optical variation behavior with the \gr\ one, 
we approximated the 
corresponding \gr\ flux at $x_{\rm b}$ for the first two types of color changes.
The method we used is as follows. For each \gr\ flux data point of a target,
we found its closest $zr$ measurement in time, but also required the former's
(30-day) time bin covering the latter's date. Thus, a plot of \gr\ flux versus
$zr$ flux density was constructed (see Figure~\ref{fig:go} as an example). 
Then based on $x_{\rm b}$ (Figure~\ref{fig:go}) 
determined above for this target, we chose two \gr\ fluxes closest
to $x_b$ in time and averaged the values 
as the corresponding \gr\ flux at the turning point.
The flux values obtained for 31 blazars of the first two types of color 
changes are provided in Table~\ref{Tab1}.

\subsection{{\it Swift} X-ray data analysis}

It is interesting to learn the variation behavior at other wavelengths
during the constant-color part. We thus searched archival X-ray data 
from {\it Swift} X-ray Telescope (XRT) observations 
during the time periods when a blazar target was in the stable
state. 
Note that the 16 blazars fitted with a constant are considered to be always
in the stable state \citep{Zha+22}.
Observations of six sources were found, and the times of
the observations for each source are provided in Table~\ref{tab:xray}.

We employed the online Swift-XRT data products generator tool\footnote{\url{https://www.swift.ac.uk/user_objects/}} to obtained the 0.3--10 keV X-ray spectra. 
This tool allows the submission of a large sample of objects, and automatically 
extract and fit the spectrum in each observation (for details about the 
online tool, see \citealt{Evans+09}). In each extraction, we set an 
absorbed power-law model to fit the spectrum, where the Galactic hydrogen 
column density $N_{\rm H}$ was automatically fixed at the value 
from \citet{Willingale+13}. In total, 71 observations of the six blazars
were analyzed, and unabsorbed fluxes and photon indices ($\Gamma_x$) were 
obtained from the observations.

\begin{table}[!ht]
        \caption[]{Times of {\it Swift}-XRT observations}
    \centering
    \begin{tabular}{lc}
    \hline
        4FGL & Times \\ \hline
	    J0028.5+2001 & 8 \\
	    J0038.2$-$2459 & 2 \\
	    J1159.5+2914 & 31 \\
	    J1310.5+3221 & 19 \\
	    J1350.8+3033 & 4 \\
	    J1700.0+6830 & 7  \\ \hline
	     \end{tabular}
	     \label{tab:xray}
\end{table}

\section{Results and Discussion} 
\label{sec:rd}

By investigating color-magnitude changes of 47 blazars that had
the largest optical variations in the time period of $\sim$MJD~58190--60200,
we have found 31 of them showing a RWB pattern and the other 16 being
around a constant of $zg-zr$. More specifically, the color changing behavior
of the 31 blazars may be described as redder-stable-when-brighter (RSWB; e.g.,
\citealt{Zha+22,Mc+24}), since they all appeared to have a turning point over
the course of brightening, after which the $zg-zr$ color
stayed around a constant. The other 16 blazars may be described as 
stable-when-brighter (SWB; \citealt{Zha+22}). These behaviors have been
observed from earlier studies of individual sources (e.g., \citealt{Gu+06};
\citealt{vil+06}; \citealt{rai+08}; \citealt{fan+18})
and more recent studies of large samples (e.g., \citealt{Ote+22,Zha+22,Mc+24}).

The RSWB pattern is considered to be mostly associated with FSRQs, which has 
been explained
with a two-component scenario (e.g., \citealt{Gu+06,Zha+22}). There is a thermal
component arising from the accretion disk and broad-line region
in an FSRQ, whose emission has a steep spectrum, and the other one from
the jet, whose emission is non-thermal with a less steep spectrum. As
the source is brightening, the latter component contributes more to
the observed emission, thus inducing the RWB behavior. After the source
reaches the point when the latter component becomes dominant, the source
enters the stable state. Thus, the SWB behavior seen in the 16 blazars
is likely because the sources were in the high-flux, stable 
state \citep{Zha+22},
although we note that their luminosities are not particularly different,
spreading from $10^{45}$ to $10^{47}$\,erg\,s$^{-1}$ along with the sources 
with variable colors
(Color\_const and Color\_var data points in Figure~\ref{fig:lum}, respectively).
These SWB sources would have their RWB part, but the fluxes would be too 
faint to be detected by ZTF, whose detection limit is $\sim$20.8\,mag.
The near-future surveys, such as 
the Legacy Survey of Space and Time (LSST; with the Vera C. Rubin
Observatory), will be able to detect their $>$20.8\,mag turning points,
by providing much deeper light-curve data.

Among the 31 RSWB blazars, only two are BL Lacs, which are PKS~0235+164 and
TXS~0404+075. The first one has been in the samples of 
\citet{Ote+22}, \citet{Zha+22}, and \citet{Mc+24}, and they all reported 
the RWB pattern for it,
although \citet{Ote+22} noted a mild bluer-when-brighter (BWB) trend
in the brightest spectra. We have not found color-pattern information
for the second source in the literature. As pointed out by \citet{Zha+22},
BL Lacs can also show the RSWB behavior, as long as the accretion disks
are sufficiently brighter (e.g., disk luminosities are probably as large as
$5\times 10^{45}$--10$^{46}$\,erg\,s$^{-1}$).

\begin{figure}{}
\centering
\includegraphics[width=0.45\textwidth]{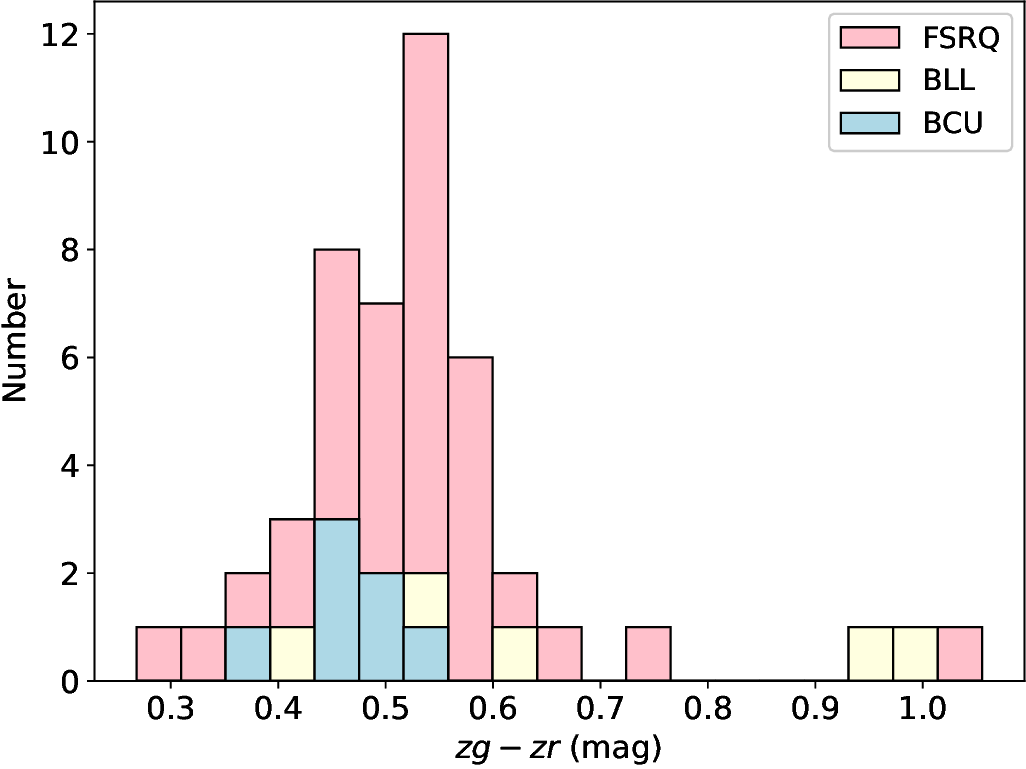}
\includegraphics[width=0.45\textwidth]{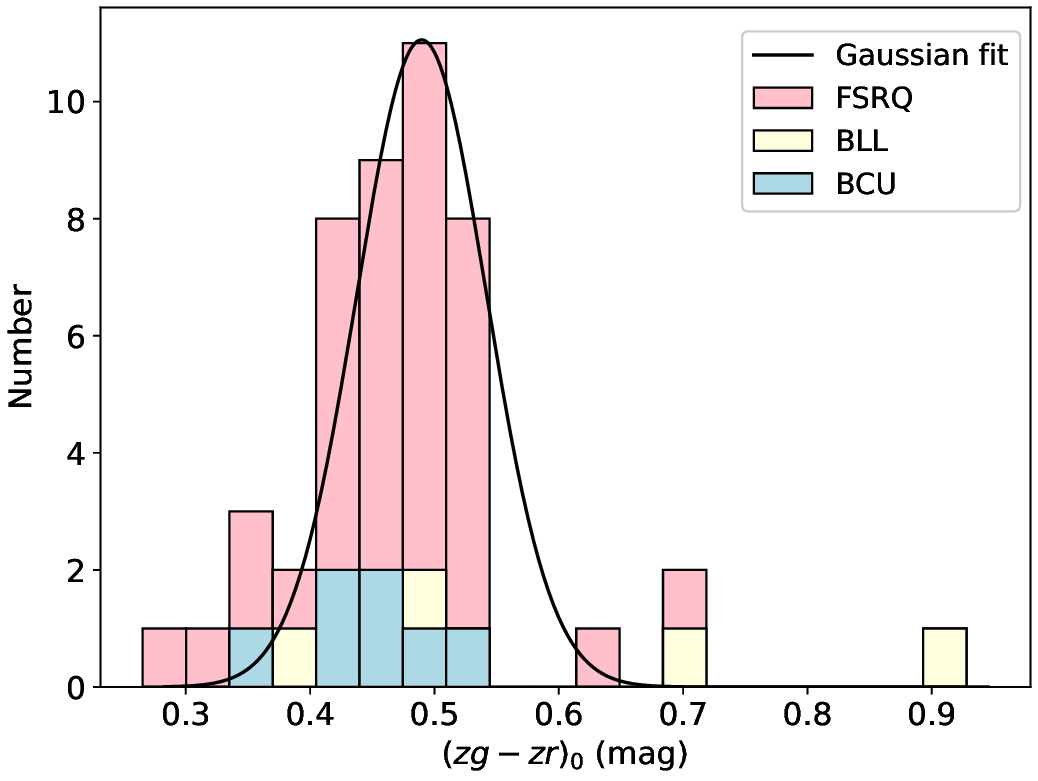}
	\caption{{\it Left:} distribution of observed $zg-zr$ at the stable
	state. {\it Right:} distribution of $(zg-zr)_0$, which has been 
	corrected with the Galactic extinction. }
 \label{fig:dist}
\end{figure}

\begin{figure}{}
\centering
\includegraphics[width=0.45\textwidth]{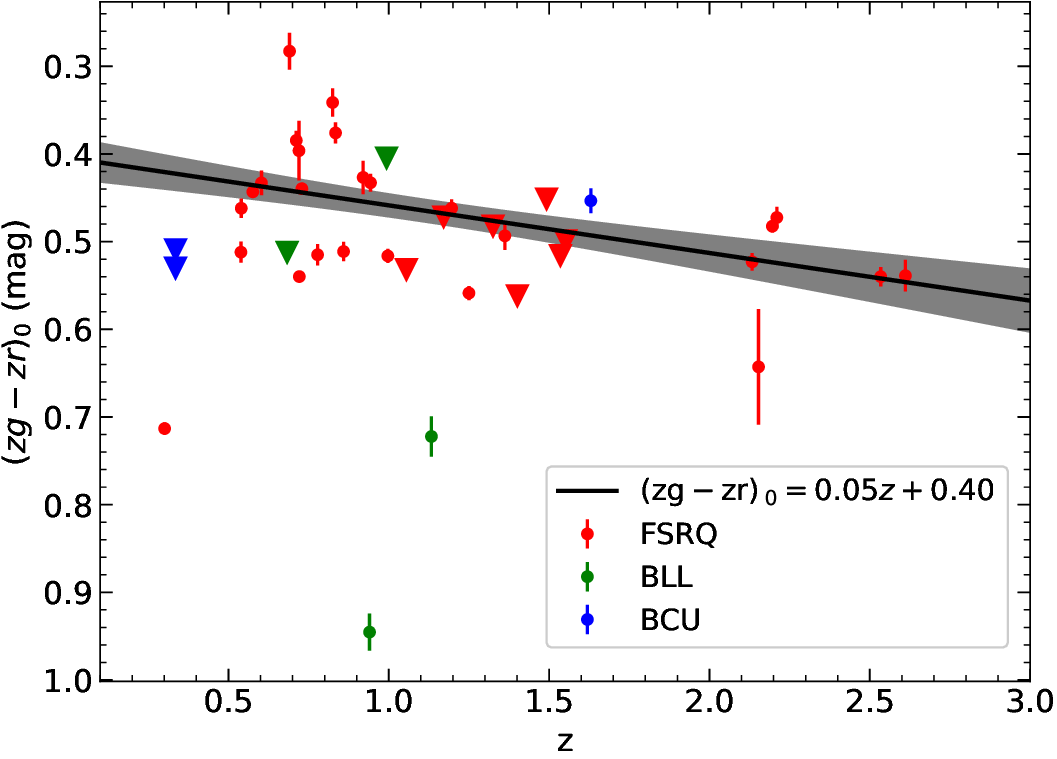}
	\caption{Dereddened color $(zg-zr)_0$ as a function of
	redshift $z$. Downward triangles mark those blazars not showing 
	a turning point in their color-magnitude diagrams. 
	A weak trend of $(zg-zr)_0 \propto 0.05z$ is possibly
	seen. }
 \label{fig:red}
\end{figure}

In our analysis, we noted that the color values in the stable state
are similar. We plot the color distribution in the left panel of 
Figure~\ref{fig:dist}. Most sources are within $\sim$0.4--0.6, but
several have larger values. We adopted the estimates of the
Galactic extinction $E(B-V)$ values to our sources from \citet{sf11}, 
and the values are provided in Table~\ref{Tab1}. We then dereddened
$zg-zr$ with the respective extinction values for the sources
(where we approximated with reddening at $zg$, $A_{zg} \simeq 3.303E(B-V)$,
and that at $zr$, $A_{zr} \simeq 2.285E(B-V)$; \citealt{sf11}).
The dereddend color $(zg-zr)_0$ distribution is shown in the right panel
of Figure~\ref{fig:dist}. As can be seen, the sources are slightly more 
concentrated to a range of 0.4--0.55. A Gaussian fit can approximately
describe the distribution, with the peak at $\simeq$0.49 and a standard 
deviation of $\simeq$ 0.052. 
There are three notable sources with 
$(zg-zr)_0> 0.7$, two of which are the BL Lacs PKS~0235+164 (at $\sim$0.9) 
and TXS~0404+075 (at $\sim$0.7). These two are the only known 
BL Lacs mentioned above that showed the RSWB pattern in our sample. 
Given this, we might consider that BL Lacs tend to have redder colors in 
the stable state. However, we note that towards PKS~0235+164 ($z=0.94$), there 
is a known absorber at redshift $z=0.524$ \citep{jun+04,rai+05}.
Due to this case, the large color values could likely be caused by 
extra-absorption along the line of sight towards the sources. 

\begin{figure}{}
\centering
\includegraphics[width=0.4\textwidth]{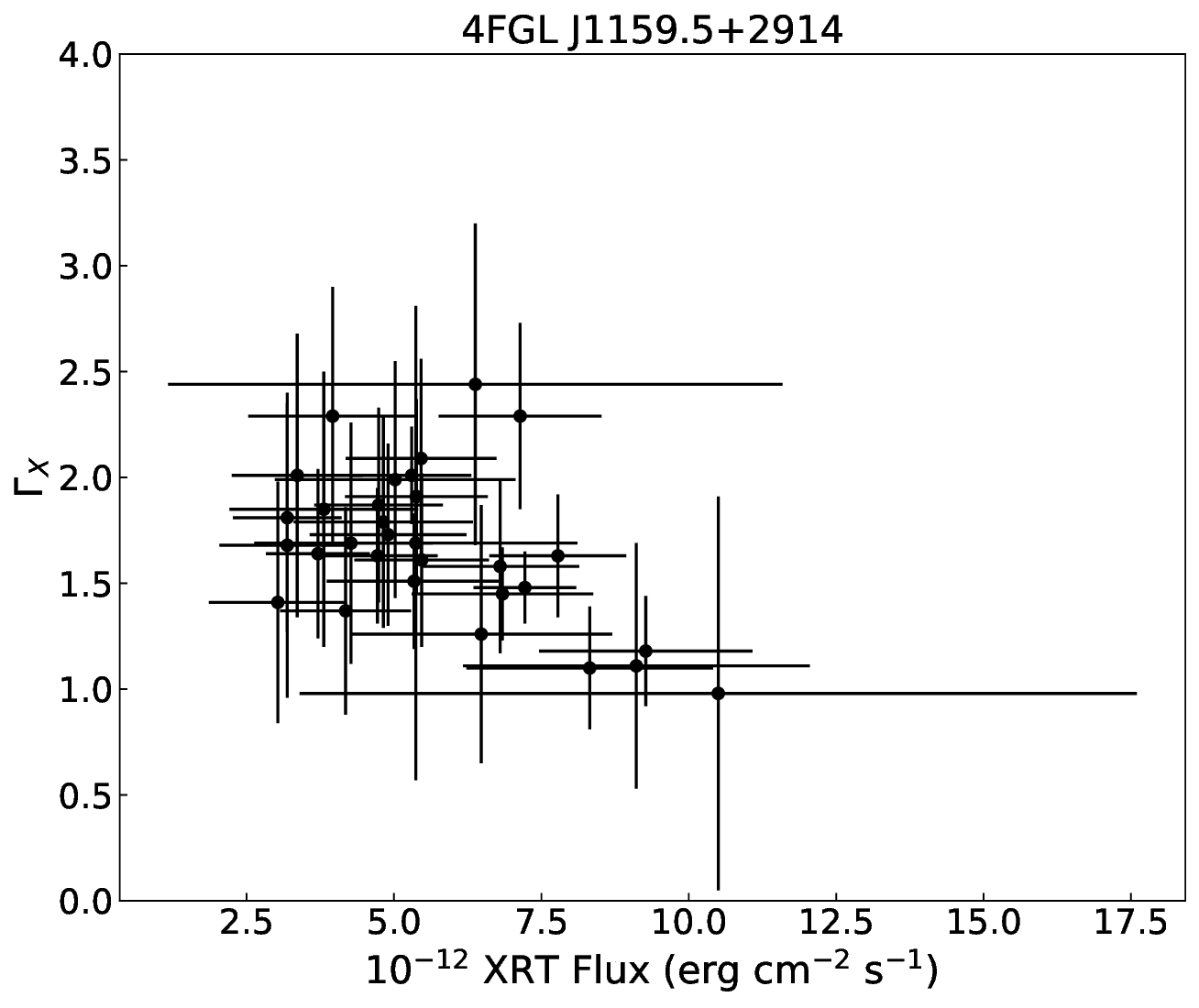}
	\caption{X-ray photon index ($\Gamma_x$) versus 
	0.3--10\,keV X-ray
	flux of Ton~599 during its optical stable state. A brighter and
	harder (lower $\Gamma_x$ values) pattern is seen.}
 \label{fig:ton}
\end{figure}


We checked $(zg-zr)_0$ as a function of $z$ with the plot shown in 
Figure~\ref{fig:red}. The redshift ranges from 0.3 to 2.6, which
shift emission at $\sim$5000$/(1+z)$\,\AA\ to the observed bands. The largest
value of $z\simeq 2.6$ corresponds to an ultraviolet (UV) wavelength 
of $\sim 1300$\,\AA.
Except the three $(zg-zr)_0 > 0.7$ sources mentioned above, the sources do 
not show large color variations or a strong trend. If we consider
a power-law spectrum $F_\nu\propto \nu^{-\alpha}$ for the optical-UV emission at
the rest wavelengths, the narrow range of $(zg-zr)_0\sim$0.4--0.55 would suggest
that there are no large changes in $\alpha$ from UV to optical in these
sources.
We tried to fit the colors as a function $z$ (not including
the three sources with $(zg-zr)_0 > 0.7$). A possible trend of 
$(zg-zr)_0 \propto 0.05 z$ was obtained, although the colors are relatively 
largely scattered around the fit (Figure~\ref{fig:red}). In any case, this
weak trend could be the indication of the extra-reddening caused by
the intragalactic medium (IGM), if we assume the UV-optical spectra of blazars
in the stable state have highly similar $\alpha$ values. In X-rays,
the IGM has been detected as the extra-absorption in spectra of
blazars (and gamma-ray bursts; see \citealt{gat+24} and references therein).
A large sample of blazars with the stable-state colors would provide a test
for this possibility (we will carry out the study in another work).

We also checked if there were significant X-ray spectral changes during
the stable state of the blazars. {\it Swift} XRT data for six
sources were available (Table~\ref{tab:xray}). Except 4FGL~J1159.5+2914 
(or Ton~599), no significant changes were detected. 
However, because
the X-ray fluxes of the sources were in a range of 
$\sim$(1--10)$\times 10^{-12}$\,erg\,cm$^{-2}$\,s$^{-1}$ and the exposures
of the {\it Swift} observations were short (generally $<$1000 sec), 
the measurements
suffer large uncertainties (see, e.g., Figure~\ref{fig:ton}). In any case,
no evidence of spectral changes at X-rays were found for the sources
at the optical stable state. In Ton~599, we detected a harder when brighter 
pattern (Figure~\ref{fig:ton}). However, its SED likely consists of
contributions from multiple components of two emission zones
in its brightening state: its X-ray emission probably arises dominantly 
from a self ICS process in an out zone with the seed photons provided 
by the jet, while the optical emission mainly comes from the synchrotron
radiation in an inner zone \citep{pc20}. The connection between the optical 
and X-ray emissions in this source is likely not direct, which
may explain the spectral variation at X-rays in the optical stable state.

\begin{figure}{}
\centering
\includegraphics[width=0.5\textwidth]{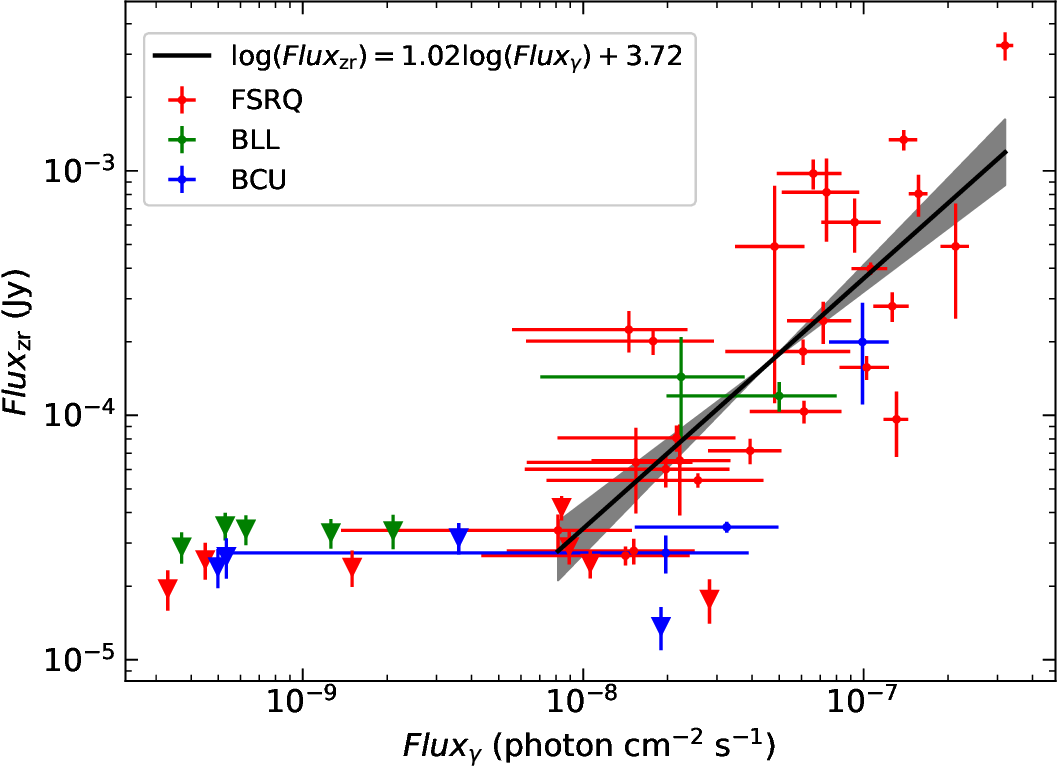}
  \caption{$zr$ flux density $F_{zr}$ versus \gr\ flux $F_\gamma$ at 
	the turning 
	points of the 31 blazars. A $F_{zr} \propto F_\gamma^{1.02}$ 
	relationship is seen.  The 16 blazars not showing a turning point,
	marked with the Downward triangles (i.e., the flux-density upper limits 
	in $zr$-band), are also included for comparison.} 
 \label{fig:fig11}
\end{figure}

As pointed out by \citet{Zha+22}, the turning point from the RWB to 
the stable
state would commonly exist, brighter than which the non-thermal, synchrotron emission
become dominant in the observed optical emission. 
For the 31 blazars showing such a turning point, we notice that
above it, the data points were in one or several flares, which often have
the corresponding \gr\ flares. The \gr\ activity possibly reflects the general
state of such a blazar is at.  In Figure \ref{fig:fig11}, we show
the $zr$-band flux $F_{zr}$ and the corresponding \gr\ flux $F_\gamma$ 
of the 31 blazars (i.e., values in the last two columns of Table~\ref{Tab1}).
A correlation between the optical and \gr\ fluxes is seen, and
$F_{zr} \propto F_\gamma^{1.02\pm0.16}$. We also investigated if there is
some physical condition for the turning point that could be indicated in
$\gamma$-rays. For example, a comparison of $F_{zr}$ versus $F_\gamma$,
normalized respectively by the maximum $zr$ flux density and the maximum
\gr\ flux of a source, 
was conducted.  However, because of large uncertainties in \gr\ fluxes
(cf., Figure~\ref{fig:fig11}),
no clear results were seen. Such investigations could be further conducted 
by selecting only bright \gr\ blazars in a sample.

As a summary, we have analyzed the ZTF optical data for 47 \fermi-LAT blazars 
that showed the largest optical variations in the past several years. Among
them, 33, 7, and 7 are FSRQs, BL Lacs, and BCUs, respectively, and all are
LSPs. Their color changes show either a RSWB pattern or a constant-line 
pattern, in the former the colors reach a constant-line 
state (or the stable state) after a turning point.
The colors at the stable state share similar values, and 
the dereddened values were obtained. Except 
three sources have large color values of $>$0.7, which could be caused by
the existence of
extra-absorbers along the line of sight towards them based on one known case
PKS 0235+164, most of the sources have the color values in a range of
0.4--0.55. If this feature of a narrow color range is true for blazars 
in the stable state, it could be used to explore the IGM. In addition, a
correlation between optical flux and \gr\ flux at the turning point is found,
but what this correlation implies remains to be investigated.

\begin{acknowledgements}
This work was based on observations obtained with the Samuel Oschin Telescope
48-inch and the 60-inch Telescope at the Palomar Observatory as part of
the Zwicky Transient Facility project. ZTF is supported by the National
Science Foundation under Grant No. AST-2034437 and a collaboration including
Caltech, IPAC, the Weizmann Institute for Science, the Oskar Klein Center
at Stockholm University, the University of Maryland, Deutsches
Elektronen-Synchrotron and Humboldt University, the TANGO Consortium of
Taiwan, the University of Wisconsin at Milwaukee, Trinity College Dublin,
Lawrence Livermore National Laboratories, and IN2P3, France. Operations are
conducted by COO, IPAC, and UW.

	We thank the referee for detailed and helpful comments.
This research is supported by the Basic Research Program of Yunnan
Province (No. 202201AS070005), the National Natural Science Foundation
of China (12273033), and the Original
Innovation Program of the Chinese Academy of Sciences (E085021002).
S.J. acknowledges the support of the science research program for graduate 
students of Yunnan University (KC-23234629).
	\end{acknowledgements}

\bibliographystyle{raa}
\bibliography{ms2024-0200}

\appendix
\section{Color Magnitude Diagrams of 44 Blazars}
\label{sec:cm}

\begin{figure*}
\includegraphics[width=0.95\textwidth]{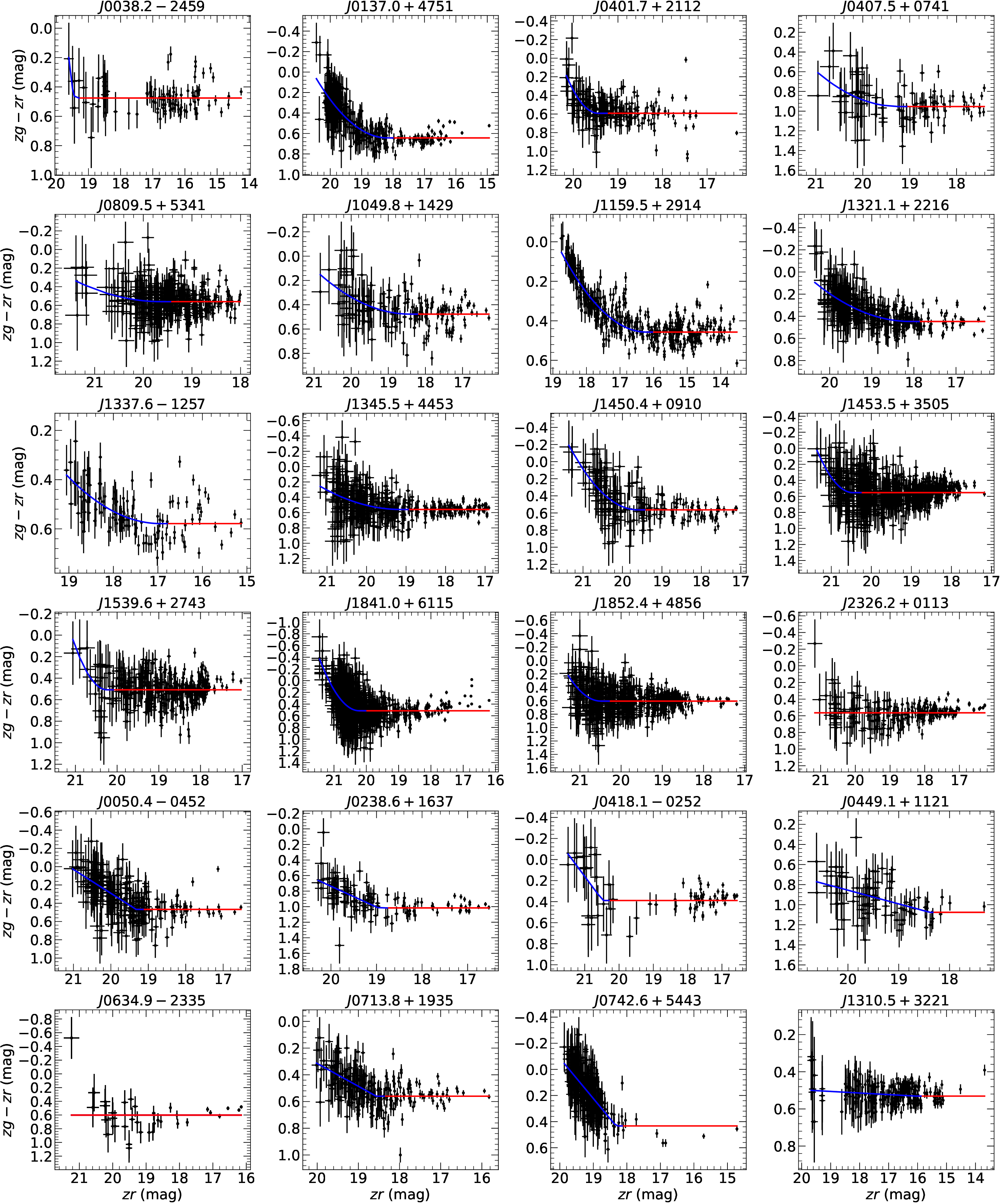}
	\caption{Color magnitude ($zg-zr$ versus $zr$) diagrams for 24 blazar
	targets. The blue line indicates the fit to the BWB part and the red
	line the fit to the stable part. The magnitudes at the turning points,
	with the uncertainties from the fitting, are given in Table~\ref{Tab1}.
	Note that the panels have different scales, which are set to include
	all the data points for each target.
	No error regions from the fitting are shown for clarity.}
	\label{fig:cm1}
\end{figure*}

\begin{figure*}
\includegraphics[width=0.95\textwidth]{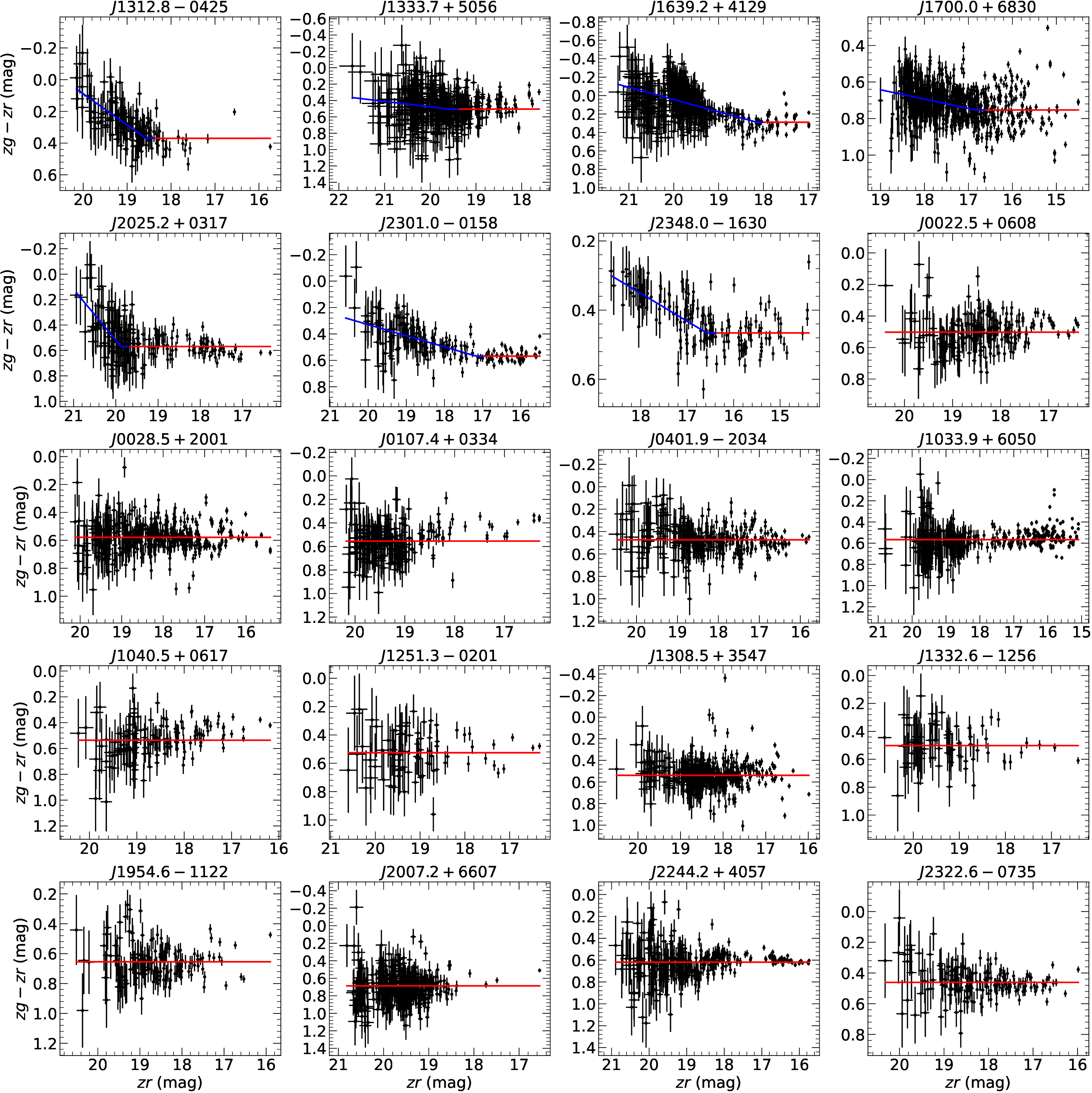}
	\caption{Same as Figure~\ref{fig:cm1} for 20 blazar targets.}
	\label{fig:cm2}
\end{figure*}

\end{document}